\documentclass[prl, reprint,amsmath,amssymb,nofootinbib,superscriptaddress,floatfix,showpacs]{revtex4-1}

\usepackage{graphicx}
\usepackage{dcolumn}
\usepackage{bm}
\usepackage{txfonts}

\DeclareMathOperator{\sech}{sech}

\begin{document}

%
%
\title{Nonreciprocal spin-wave channeling along textures driven by the Dzyaloshinskii-Moriya interaction}

\author{Felipe Garcia-Sanchez}
\affiliation{Institut d'Electronique Fondamentale, Univ. Paris-Sud, 91405 Orsay, France}
\affiliation{UMR 8622, CNRS, 91405 Orsay, France}
\author{Pablo Borys}
\affiliation{SUPA, School of Physics and Astronomy, University of Glasgow, Glasgow G12 8QQ, UK}
\author{Arne Vansteenkiste}
\affiliation{Department of Solid State Sciences, Ghent University, Krijgslaan 281-S1, B-9000 Ghent, Belgium}
\author{Joo-Von Kim}
\email{joo-von.kim@u-psud.fr}
\affiliation{Institut d'Electronique Fondamentale, Univ. Paris-Sud, 91405 Orsay, France}
\affiliation{UMR 8622, CNRS, 91405 Orsay, France}
\author{Robert L. Stamps}
\affiliation{SUPA, School of Physics and Astronomy, University of Glasgow, Glasgow G12 8QQ, UK}

\date{\today}

\begin{abstract}
Ultrathin metallic ferromagnets on substrates with strong spin-orbit coupling can exhibit induced chiral interactions of the Dzyaloshinskii-Moriya (DM) form. For systems with perpendicular anisotropy, the presence of DM interactions has important consequences for current-driven domain-wall motion and underpins possible spintronic applications involving skyrmions. We show theoretically how spin textures driven by the DM interaction allow nonreciprocal channeling of spin waves, leading to measurable features in magnetic wires, dots, and domain walls. Our results provide methods for detecting induced DM interactions in metallic multilayers and controlling spin wave propagation in ultrathin nanostructures.
\end{abstract}


\maketitle

The Dzyaloshinskii-Moriya interaction (DMI) has been used to explain canted states in weak ferromagnets and antiferromagnets and can appear when crystal structure allows or structural defects exist in such a way as to remove inversion symmetry~\cite{Dzyaloshinsky:1958vq, Moriya:1960go, Moriya:1960kc}.  Some weak ferromagnets also display multiferroicity with simultaneous magnetic- and electric-field response, and the DMI can be associated with magnetoelectric interactions~\cite{Ederer:2005hk, Ederer:2008cy, Varga:2009cu}. A class of systems admit fascinating chiral spin textures described in terms of DMI, including helicoidal and skyrmionic~\cite{Muhlbauer:2009bc, Yu:2010iu, Heinze:2011ic, Seki:2012ie} orderings. Skyrmions in particular have attracted much recent attention for spintronics as a result of their unique properties involving propagation under spin polarized currents, such as dynamics under ultralow critical current densities~\cite{Jonietz:2010fy} and high tolerance to material defects~\cite{Iwasaki:2013ji, Sampaio:2013kn}. Experiments illustrating the controlled nucleation and annihilation of individual skyrmions pave the way towards new applications for information storage and processing~\cite{Romming:2013iq}.

Ultrathin films lack inversion symmetry simply because they are grown on a substrate of one material and are possibly capped with a different material. The resulting structure by definition has asymmetric interfaces and therefore also falls into this class of low symmetry structure regardless of its underlying atomic symmetry. In such films there is then at least the possibility of a DMI~\cite{Bogdanov:2001hr, Rossler:2006cq}. One microscopic mechanism for interface-driven DMI involves the presence of significant spin-orbit coupling at one interface of the ultrathin film. Experiments have shown that such induced chiral interactions can lead to modulated chiral spin structures in manganese monolayers on tungsten~\cite{Bode:2007em} and skyrmion lattices in iron monolayers on iridium~\cite{Heinze:2011ic}, where the spin configurations observed are in good agreement with electronic structure calculations. For other candidate systems, such as Pt/Co, it has been argued that a three-site indirect exchange mechanism should lead to an interfacial chiral interaction in Co of the Dzyaloshinskii-Moriya form~\cite{Fert:1980hr}, with the same symmetry expected for films with perpendicular anisotropy~\cite{Bogdanov:2001hr}. It is therefore an intriguing prospect to consider that strong chiral interactions may have been present, but unrecognized, in materials of key interest for spintronics: namely, Pt/Co systems with perpendicular magnetic anisotropy that have been studied for over a decade for possible applications in magnetic storage.

At present, evidence of the DMI in systems like Pt/Co has been inferred from measurements of domain-wall dynamics~\cite{Emori:2013cl, Ryu:2013dl} and spin-polarized low-electron electron microscopy studies of static domain-wall profiles~\cite{Chen:2013bc}. In these systems, it is argued that the DMI can lead to a N{\'e}el domain-wall profile at equilibrium~\cite{Heide:2008da}. The N{\'e}el wall profile is significant in terms of its effect on wall mobilities because these walls are narrow and can travel with high velocities under applied fields or currents involving the spin Hall effect~\cite{Thiaville:2012ia}. However, direct schemes for quantifying the DMI in such multilayered structures are still lacking. While surface spectroscopy techniques allow the DMI to be determined through measurements of the asymmetric spin-wave dispersion~\cite{Udvardi:2009fm, Zakeri:2010ki}, they are less useful for nanostructures in which films are buried.

A challenging problem is therefore to measure the strength of the DMI in these ferromagnetic metals. Here, we show how the DMI may be detected and quantified through the nonreciprocal propagation of spin waves that are channeled by chiral spin textures that appear as a result of it. In particular, we show how nonreciprocity appears for N{\'e}el domain walls and how spin-wave channeling occurs at edges of wires and dots where partial walls describe local tilting in the magnetization. These effects give rise to measurable features in the spin-wave spectra of domain walls, as well as wires and dots that are nominally uniformly magnetized.

To see more clearly how the symmetry of allowed interactions control dynamic states, we show first how the DMI modifies spin textures. As discussed elsewhere~\cite{Heide:2008da, Thiaville:2012ia}, the DMI strongly modifies the profile of a domain wall by changing the sense of rotation of the spins through the wall. In perpendicular anisotropy films such a wall has a characteristic size $\lambda = \sqrt{A/K_0}$, which arises from the competition between an isotropic exchange interaction, $U_{ex} = \int dV \; A \left( \nabla \mathbf{m} \right)^2$, where $A$ is the exchange constant, and a uniaxial anisotropy along the $z$ axis normal to the film plane, $U_{K} = -\int dV \; K_0 m_z^2$, where $K_0 = K_u - \mu_0 M_s^2/2$, $K_u$ is the interface-driven uniaxial anisotropy energy, and $M_s$ is the saturation magnetization. Here, $\mathbf{m} =  \mathbf{m}(\mathbf{x},t)$ is a unit vector representing the time and spatially varying spin profile in a continuum approximation. The DMI is included by an additional term of the form~\cite{Bogdanov:2001hr, Thiaville:2012ia}
\begin{equation}
U_{\rm DM} = \int dV \; D \left[ m_z \left( \mathbf{\nabla}\cdot \mathbf{m} \right)  - \left(\mathbf{m} \cdot \mathbf{\nabla} \right) m_z \right],
\label{eq:DMI}
\end{equation}
where $D$ is the Dzyaloshinskii-Moriya constant. The form of the DMI in (\ref{eq:DMI}) leads to a preference for N\'eel domain walls over Bloch profiles~\cite{Heide:2008da, Thiaville:2012ia}.

Twisted spin states are also expected at edges with the DMI. To appreciate this, it is useful to recall that the variational procedure leading the to the torque equation, 
\begin{equation}
\frac{\partial \mathbf{m}}{\partial t} = -|\gamma_0| \mathbf{m} \times \left( -\frac{1}{\mu_0 M_s} \frac{\delta U}{\delta \mathbf{m}}  \right),
\label{eq:torque}
\end{equation}
where $U = \int dV \; \mathcal{U}$ is the total energy, also gives rise to a boundary condition of the form, $\mathbf{n} \cdot \partial \mathcal{U}/\partial (\nabla \mathbf{m})   = 0$, where $\mathbf{n}$ is a unit vector normal to the surface of the material considered~\cite{SM}. With only $U = U_{ex} + U_K$, one obtains the usual free boundary condition, $\partial_{\mathbf{n}} \mathbf{m} = \mathbf{0}$, in the absence of any surface pinning. Crucially, the inclusion of Eq.~\ref{eq:DMI} in $U$ requires satisfaction of twisted boundary conditions. For example, the boundary surface $\mathbf{n} = \hat{\mathbf{y}}$ has the conditions
\begin{equation}
D m_z + 2 A \,\partial_y m_y = 0; \;\; -D m_y + 2 A \, \partial_y m_z = 0,
\label{eq:bc}
\end{equation}
which couples the perpendicular magnetization $m_z$ with gradients in the transverse components $m_{x,y}$, and vice versa~\cite{SM, Rohart:2013ef}. Such conditions lead to tilts in the magnetization at the edges even if the system is uniformly magnetized in the bulk.

\begin{figure}
\centering\includegraphics[width=8.5cm]{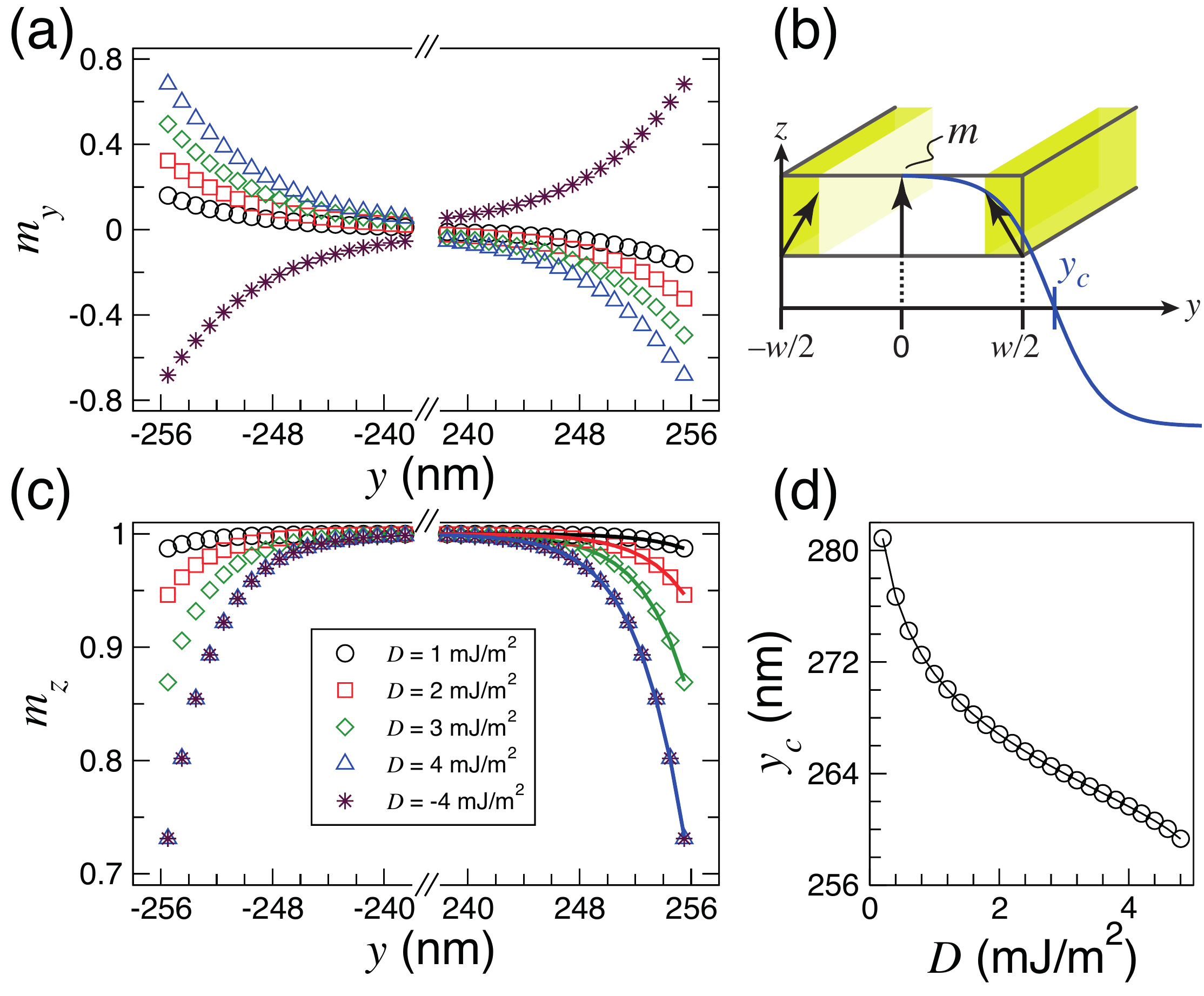}
\caption{(Color online) (a) The transverse magnetization component $m_y$ at the boundary edges (located at $y = \pm256$ nm) of a 512 nm wide rectangular wire. (b) Illustration of the magnetization tilts for $D>0$, with the yellow shaded regions representing the tilts shown in panels (a) and (c). The partial wall (blue curve) is shown schematically, with $y_c$ denoting the wall center and $w$ the wire width. (c) The perpendicular  component $m_z$  at the boundary edges, where the solid lines correspond to fits to a partial N{\'e}el wall profile. (d) Partial wall center $y_c$ as a function of $D$.}
\label{fig:tilts}
\end{figure}

An example of magnetization tilts at edges is shown in Fig.~\ref{fig:tilts}. 
The profiles were computed with micromagnetics simulations~\cite{Vansteenkiste:2011bx, mumax3} by first allowing a uniformly magnetized state in a $512$ nm $\times$ $512$ nm $\times$ 1 nm square dot to relax under several values of the DMI~\cite{SM}. Stronger tilts occur when the strength of the DMI is increased, and the sign of the transverse component of the tilts is reversed along with the sign of the DMI [Figs.~\ref{fig:tilts}(a) and ~\ref{fig:tilts}(b)]. These profiles are well described by \emph{partially expelled} N{\'e}el walls. Examples are shown by the solid curves in Fig.~\ref{fig:tilts}(c), which represent the theoretical wall profile $m_z(y) = \tan\left[(-y-y_c)/\lambda\right]$ at the right edge, where $y_c$ is the position of the domain-wall center \emph{outside} the film, as illustrated schematically in Fig.~\ref{fig:tilts}(b). This behavior is reminiscent of the partial twists encountered in exchange spring systems and ferromagnet/antiferromagnet bilayers where the gradual rotation of the uniformly magnetized hard (ferromagnetic) layer creates torques at the interface that are compensated by formation of a partial wall structure in the soft (antiferromagnetic) layer~\cite{Suess:2005, GarciaSanchez:2006, Mauri:1987vt, Kim:2005jf}. Here, the DMI acts to pin a partial wall at the edges through Eq.~\ref{eq:bc}, and the strength of the DMI governs the extent to which the partial wall enters the film [Fig.~\ref{fig:tilts}(d)].

Dynamic collective excitations above this tilted ground state are spin waves, which can be described by  equations of motion in the low energy, long wavelength limit by linearizing (\ref{eq:torque}) with $\mathbf{m}(\mathbf{x},t) = \mathbf{m}_0(\mathbf{x}) + \delta \mathbf{m}(\mathbf{x},t)$, where $\mathbf{m}_0(\mathbf{x})$ describes the static configuration and $\delta \mathbf{m}(\mathbf{x},t)$ represents the spin-wave fluctuations. Moon \emph{et al}. have shown that inclusion of a DM term into $U$ allows a term linear in the spin-wave propagation vector in the case of a uniformly magnetized, infinitely extended planar film,  thereby creating nonreciprocity (i.e., $\omega(k) \ne \omega(-k)$ for some propagation directions)~\cite{Moon:2013dm}. Similar results were reported earlier for monolayer Fe films for higher-energy excitations~\cite{Udvardi:2009fm, Zakeri:2010ki}.

Chiral interactions also create nonreciprocity for spin-wave propagation along the edges of magnetic wires and dots. We can understand how the DMI-induced edge texture affects spin-wave propagation by examining propagation across and along a one-dimensional domain boundary wall. With only $U = U_{ex} + U_K$, the domain wall appears as a reflectionless potential for spin waves traveling across the wall axis~\cite{Winter:1961hw}. In this case, one possible process involving a static wall is an acquired phase that accompanies the complete transmission of the spin-wave through the domain wall~\cite{Hertel:2004df, Bayer:2005ev}. However, the DMI deforms the wall profile such that the potential is no longer reflectionless, and a traveling spin-wave hybridizes with wall localized states and is partially reflected. Travel along the wall axis is different. Propagation in this direction also requires the spin-wave to be in one of the hybridized states but positive or negative wall axis directions are not equivalent when the DMI is present, resulting in spin-wave channels for right ($+x$) and left ($-x$) propagation that have different energies.

\begin{figure}
\centering\includegraphics[width=8.5cm]{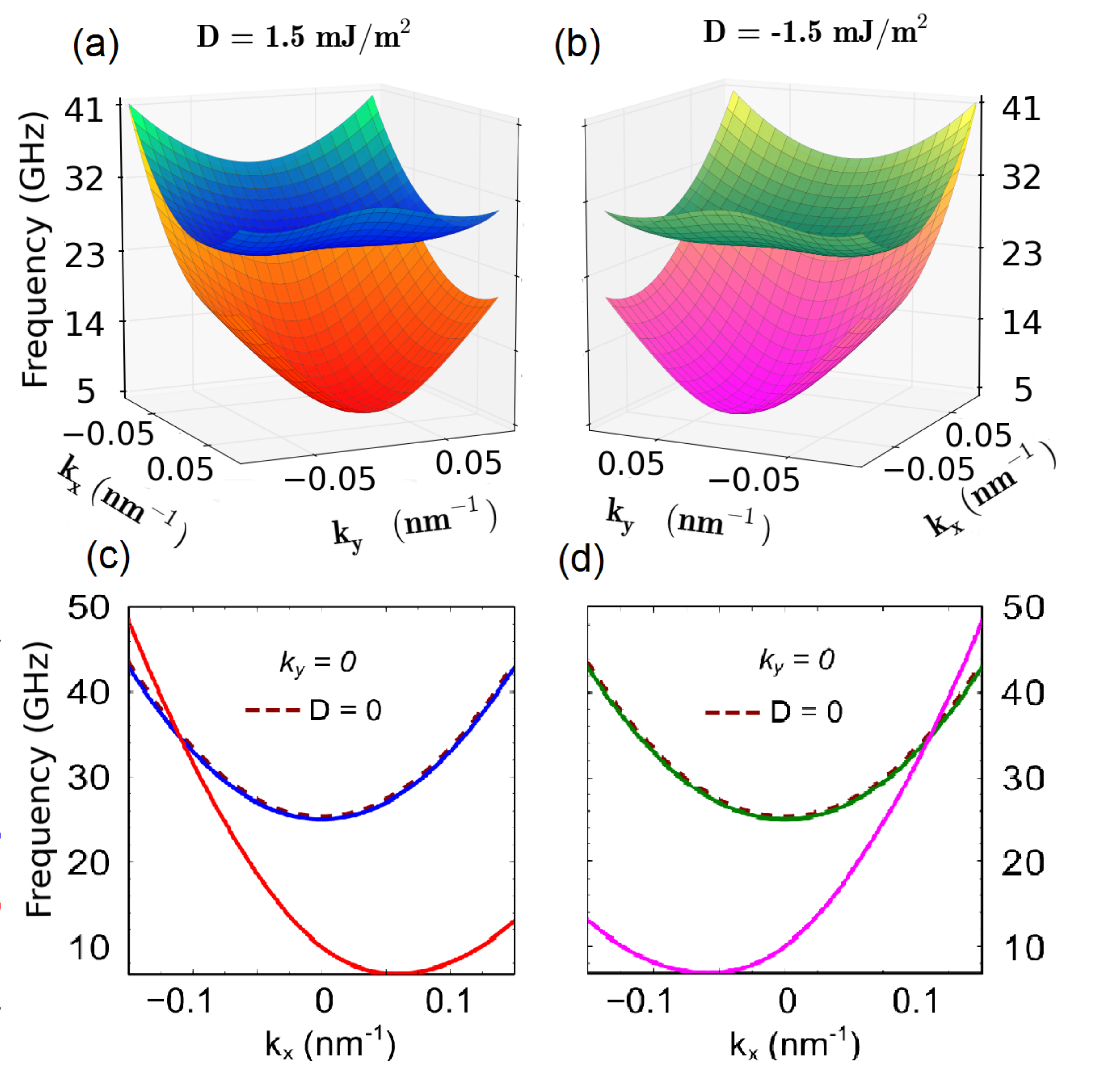}
\caption{(Color online) N{\'e}el wall eigenfrequencies calculated using perturbation theory for weak DMI. (a), (b) The DMI splits frequencies into two sheets that are otherwise degenerate and distorts the sheets such that propagation is nonreciprocal with respect to $k_x$  $\left[ \omega(k_x)\ne\omega(-k_x) \right]$. Dispersion relations for  states propagating along the domain wall ($k_y = 0$) for (c) $D = 1.5$ mJ/m$^2$ and (d) $D = -1.5$ mJ/m$^2$.}
\label{fig:3d}
\end{figure}

Degenerate-state perturbation theory is required to quantify the degree of nonreciprocity introduced by the DMI. By treating the case where the DMI is weak compared to the isotropic exchange, Eq.~(\ref{eq:torque}) can be solved for linear spin-wave propagation at arbitrary directions with respect to the wall axis~\cite{SM}. Example results are shown in Fig.~\ref{fig:3d} where propagation across ($y$ direction) and along ($x$ direction) the wall  are contrasted. The DMI lifts the degeneracy between propagating states that exists when $D=0$.  In Fig.~\ref{fig:3d}(a), one sees that counterpropagating states at a given $k$ have different frequencies when propagating along the wall ($k_y=0$). Propagation along a wall with opposite chirality is shown in Fig.~\ref{fig:3d}(b), and we see that chirality controls the nonreciprocity of the propagation. The general structure of the dispersion is shown in Fig.~\ref{fig:3d}(c) for the same chirality as in Fig.~\ref{fig:3d}(a). 

The consequences for propagation along the edges of the spin texture induced by the DMI now follow.  As discussed previously, domain walls are pinned outside any finite-sized thin-film element, but the tail of the walls remain and have the same chirality. As a result, the energies of spin-wave states propagating along a given edge will depend on their propagation direction due to the asymmetry introduced by the DMI for the $k_y = 0$ states [Figs.~\ref{fig:3d}(a) and ~\ref{fig:3d}(b)]. In consequence, the lowest energy spin waves propagate only along one direction when localized on one side of the wire and flow in the opposite direction when localized on the other side.

\begin{figure}
\centering\includegraphics[width=8.5cm]{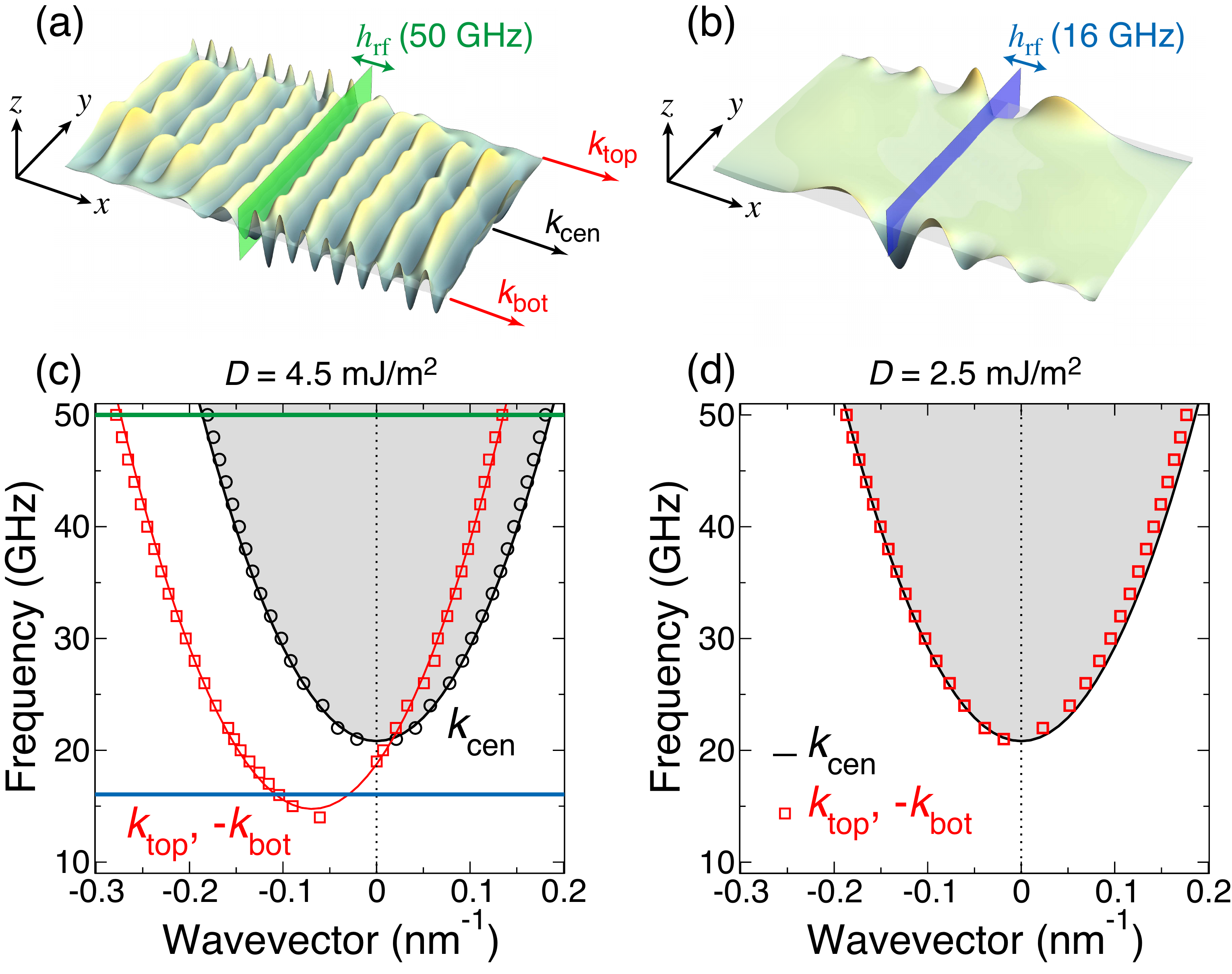}
\caption{\label{fig:wire_propagation}(Color online) Nonreciprocal propagation in a thin rectangular wire. Spatial profiles of $m_x$ resulting from a rf field excitation, $\mathbf{h}_{\rm rf}(t) = h_0 \sin(2 \pi f_{\rm rf} t) \hat{\mathbf{x}}$, where $\mu_0 h_{0} = 5$ mT, at (a)  $f_{\rm rf} = 50$ GHz and (b) $f_{\rm rf} = 16$ GHz. The different wave-vector components considered are illustrated. In panel (a), $f_{\rm rf}$ is in the spin-wave band and nonreciprocal propagation occurs for $k_{\rm top}$ and $k_{\rm bot}$, while $k_{\rm cen}$ propagation is symmetric.  In panel (b), $f_{\rm rf}$ is in the gap of the bulk modes and only edge modes are excited. (c) Dispersion relations computed from simulations for $D_{\rm ex} = 4.5$ mJ/m$^2$, with $f_{\rm rf}$ used in panels (a) and (b) indicated. Dots represent simulation results. The solid black curve (and gray shaded area) represents the theoretical dispersion relation for exchange modes. The solid red curve represents  the fit given by Eq.~(\ref{eq:disprelshifted}). (d) Dispersion relation for $D_{\rm ex} = 2.5$ mJ/m$^2$. }
\end{figure}

To examine this nonreciprocal propagation in detail, we performed micromagnetic simulations of spin-wave propagation in a thin rectangular wire~\cite{SM}.  An example of the spin waves found for a 256-nm-wide wire is given in Fig.~\ref{fig:wire_propagation}.
A pulsed magnetic field, with a spatial extension of 1 nm, was applied across the width and at the center of the 2048-nm-long rectangular stripe and the wave vector of the excited spin waves for different excitation frequencies was computed. From this analysis, the dispersion relation for propagating edge and bulk spin waves for different strengths of the DMI was constructed. For excitation frequencies in the spin-wave band [Fig.~\ref{fig:wire_propagation}(a)], $f_{\rm rf} = 50$ GHz, three distinct wave vectors can be identified for propagation along one direction, which correspond to the top ($k_{\rm top}$), center ($k_{\rm cen}$), and bottom ($k_{\rm bot}$) of the wire. For propagation towards the right, $+x$, we note that $|k_{\rm top}| < |k_{\rm cen}| < |k_{\rm bot}|$, while for propagation towards the left, $-x$, the opposite inequality applies, $|k_{\rm top}| > |k_{\rm cen}| > |k_{\rm bot}|$. Moreover, $k_{\rm top} = -k_{\rm bot}$, which is a clear signature of nonreciprocal propagation. We observe a shifted quadratic dispersion relation for the edge modes, while the central modes remain symmetric about $k_{\rm cen} = 0$ [Fig.~\ref{fig:wire_propagation}(c)]. For the central modes $k_{\rm cen}$, the dispersion relation is well described by exchange-dominated spin waves, where the theoretical curve using our micromagnetic parameters, $\omega = (2\gamma/M_s)\left( A k_{\rm cen}^2 + K_0 \right)$, agrees well with the simulated curves. For the edge modes, the shifted dispersion relation for $D = 4.5$ mJ/m$^2$ is well described by the fit [solid red line in Fig.~\ref{fig:wire_propagation}(c)]
\begin{equation}
\omega = \frac{2\gamma}{M_s}\left( A k_{\rm top}^2 + 0.9 K_0 + 0.46 D k_{\rm top} \right).
\label{eq:disprelshifted}
\end{equation}
This describes a reduction in the spin-wave gap $K_0$ due to the reduced anisotropy field at the edge in addition to a linear wave-vector term that describes the nonreciprocity. As Fig.~\ref{fig:tilts}(d) shows, the center of the partial wall is located farther outside for smaller values of the DMI, which results in a weaker nonreciprocal channeling effect. This can be seen in the dispersion relation of the edge modes in Fig.~\ref{fig:wire_propagation}(d), where the shifts become less pronounced as $D$ decreases.

Channeling as demonstrated for the wire geometry is robust with regards to the curvature of the edge. In a circular dot, for example, it is known that clockwise (CW) and counterclockwise (CCW) propagating azimuthal spin waves are degenerate in frequency. The inclusion of the DMI, however, lifts this degeneracy by favoring one handedness over the other. To appreciate how this might occur, one can imagine the edge modes in a circular dot constructed by deforming a rectangular wire bent into a ring-shaped structure.  The lowest frequency spin waves traveling along outer circumference can propagate with only one handedness. Spin waves traveling along the inner circumference travel with the opposite handedness at the same frequency.

\begin{figure}
\centering \includegraphics[width=8.5cm]{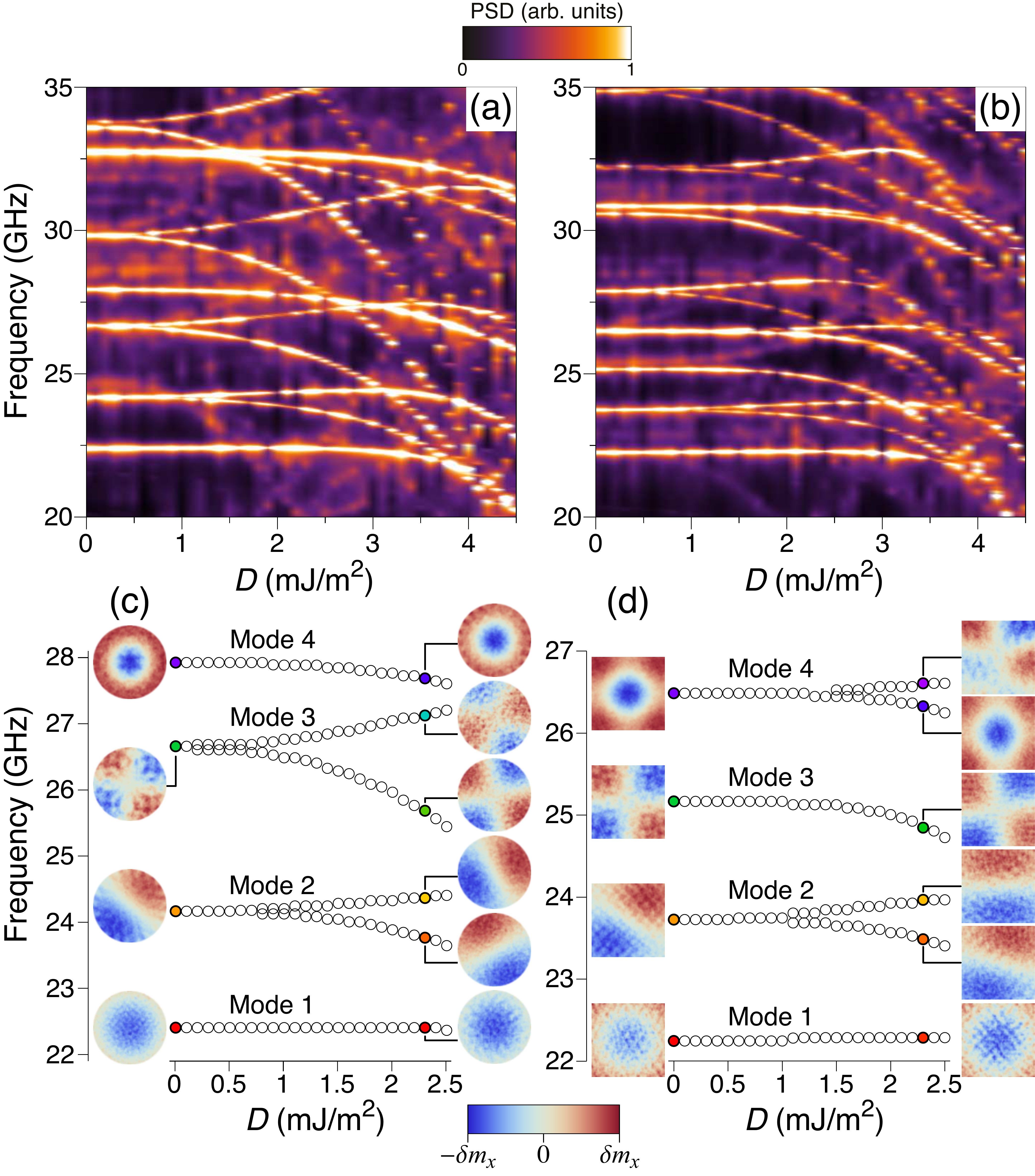}
\caption{(Color online) Map of the eigenmode power spectral density (PSD) as a function of $D$ for (a) 100-nm-diameter circular dots and (b) 100-nm-wide square dots. Selected profiles of the four lowest modes for different strengths of the DMI for the (c) circular and (d) square dots.}
\label{fig:dot_eigenmodes}
\end{figure}

Figure~\ref{fig:dot_eigenmodes} illustrates the spin-wave eigenmode spectra for a circular dot 100 nm in diameter and a square dot 100 nm in width.
 A key feature is the frequency splitting of certain modes as the strength of the DMI is increased. The frequency of other modes, on the other hand, are only slightly affected by the DMI. For a similar dot size, the magnitude of the splitting appears to be larger for the circular dots, which suggests that the azimuthal component of the eigenmodes plays an important role. For the circular dots, the frequency splitting with increasing DMI is associated with lifting in the degeneracy of eigenmodes with a strong azimuthal character, such as Modes 2 and 3 in Fig.~\ref{fig:dot_eigenmodes}(c).  While there is no discernible change in the spatial profile of these modes, a frequency splitting of around 1 GHz appears at $D=2.5$ mJ/m$^2$. Modes with a strong radial character, such as Modes 1 and 4 in Fig.~\ref{fig:dot_eigenmodes}(c), experience only a slight decrease in their frequency with increasing $D$ and little change in their spatial profile. These differences can be understood in terms of the nonreciprocal wall channeling described earlier, where radial modes are similar to the $k_y \neq  0$ case for the domain-wall eigenmodes, while azimuthal modes are similar to the $k_x \neq 0$ case, which are strongly nonreciprocal. Similar features are also seen in the square dots, but the distinction between ``radial'' and ``azimuthal'' modes is not as sharp. One difference can be seen in Mode 4 in Fig.~\ref{fig:dot_eigenmodes}(d), which represents a mixed radial-azimuthal excitation for which splitting due to the DMI results in an asymmetric profile at higher frequencies.

In conclusion, we have shown theoretically that N{\'e}el domain walls driven by the Dzyaloshinskii-Moriya interaction can modify spin-wave propagation by inducing nonreciprocal channeling along the center of the wall. The channeling also occurs at the edges of wires and dots, where partial walls appear as a result of twisted boundary conditions. In dots, the DMI leads to large frequency splitting of eigenmodes with a strong azimuthal character. These features offer a means of quantifying experimentally the DMI in metallic multilayer systems relevant for spintronics.

\section{Acknowledgements}
The authors acknowledge fruitful discussions with M. Bailleul, A. Thiaville, S. Rohart, J. Sampaio, and V. Cros. This work was supported by the University of Glasgow, EPSRC, the French National Research Agency (ANR) under Contract No. ANR-11-BS10-0003 (NanoSWITI), the National Council of Science and Technology of Mexico (CONACyT), and the Flanders Research Foundation (FWO).

\newpage

\section{Supplementary Material}

\subsection{Micromagnetics simulations}
We used a modified version of the MuMax$^2$ code~\cite{Vansteenkiste:2011bx} to compute the static and dynamic magnetization states of the rectangular wires and dots. The DM-specific modifications are publicly available through MuMax$^3$~\cite{mumax3}. The code discretizes the magnetization field using the method of finite differences and performs a time-integration of the Landau-Lifshitz-Gilbert equation of motion for the magnetization dynamics,
\begin{equation}
\frac{\partial \mathbf{m}}{\partial t}  = -|\gamma_0| \mathbf{m} \times \mathbf{H}_{\rm eff} + \alpha \mathbf{m} \times \frac{\partial \mathbf{m}}{\partial t},
\end{equation}
where $\gamma_0$ is the gyromagnetic ratio, $\mathbf{H}_{\rm eff}$ is the local effective field, and $\alpha$ is the Gilbert damping constant. The material parameters used are for a model perpendicular anisotropy system: $A = 15$ pJ/m, $K_u$ = 1 MJ/m$^3$, and $M_s$ = 1 MA/m. The film thickness for all cases studied is 1 nm, meshed with one finite difference cell. For the rectangular wire (Figs. 1 and 3 of the article), the cell size in the film plane was 1 nm $\times$ 1 nm. For the calculations involving the circular and square dots (Fig. 4 of the article), the cell size in the film plane was 1.5625 nm $\times$ 1.5625 nm. Note that the characteristic wall length with these material parameters is $\lambda \approx 6.35$ nm, which is well above the discretization size used. To compute the static equilibrium magnetization configuration, a large damping constant of $\alpha = 0.5$ was taken in order to accelerate computations by working in the overdamped limit.  For the calculation of the dynamics (Figs. 3 and 4), a smaller underdamped value of the damping constant was used $\alpha = 0.001$ in order to allow long lived spin-wave modes to be distinguished.

For propagation along the wire (Fig. 3 of the article), the ground state with the DM interaction was first computed by allowing the system to relax from a uniform state. The spin waves were then generated by a sinusoidal field that was applied across the entire width of the wire, along the $y$ direction, in a region one finite difference cell wide. For each value of the sinusoidal field frequency, the resulting wave vectors were computed by performing spatial Fourier transforms of the spin-wave profiles along the $x$ direction at the top, center, and bottom of the wire.

The calculation of the dot eigenmode spectra in Fig. 4 of the article and were done in the following way. First, the zero-temperature equilibrium micromagnetic configuration was computed by allowing the initial uniform magnetization in the dot, along the $+z$ direction, to relax for 5 ns with strong damping such that the appropriate edge tilts in the magnetization were obtained. Second, a random thermal field corresponding to a temperature of 100 K was applied for 0.1 ns to introduce a small degree of nonuniformity in the magnetic configuration. Third, a Gaussian magnetic field pulse  of 5 mT in amplitude and 200 ps in width was applied in the dot plane. The resulting spin-wave spectra were then obtained from a Fourier transform of the transient response over 25 ns of the transverse component of the magnetization, which was obtained by subtracting out the zero-temperature equilibrium state.

\subsection{Exchange-DM boundary conditions}
The magnetization dynamics in the continuum approximation can be derived using a Lagrangian formalism. Consider the system Lagrangian defined as 
\begin{equation}
L = \int_\Omega dV \; \mathcal{L} + \int_\Gamma dS \; \mathcal{U}',
\end{equation}
where the first term on the right hand side contains the usual  Berry-phase term $\mathcal{T}$ and potential energy $\mathcal{U}$ contributions in the volume $\Omega$, i.e., $\mathcal{L} = \mathcal{T} - \mathcal{U}$, while the second term describes potential energy terms associated with the surface $\Gamma$ that encloses $\Omega$. For conservative dynamics, the Euler-Lagrange equations resulting from the usual variational problem leads to the well-known dynamical system for the volume magnetization,
\begin{equation}
\frac{d}{dt} \frac{\partial \mathcal{L}}{\partial \dot{\mathbf{m}}} - \frac{\delta \mathcal{L}}{\delta \mathbf{m}},
\end{equation}
where 
$\mathcal{L}$ is the system Lagrangian and $\mathbf{m} = \mathbf{m}(\mathbf{r},t)$ is a unit vector representing the time- and space-dependent magnetization field. This results in the usual torque equation for magnetization,
\begin{equation}
\frac{\partial \mathbf{m}}{\partial t} = -|\gamma_0| \mathbf{m} \times \mathbf{H}_{\rm eff},
\end{equation}
where
\begin{equation}
\mathbf{H}_{\rm eff} = -\frac{1}{\mu_0 M_s} \frac{\delta U}{\delta \mathbf{m}}
\end{equation}
is the effective field around which the magnetization precesses. However, the variational procedure also leads to boundary condition,
\begin{equation}
\frac{\partial \mathcal{U}'}{\partial \mathbf{m}} + \frac{\partial \mathcal{U}}{\partial (\nabla \mathbf{m})} \cdot \mathbf{n} = 0,
\end{equation}
where $\mathbf{n}$ is a unit vector normal to the surface $\Gamma$. In the absence of any surface terms, such as surface or interface anisotropies, $\mathcal{U}' = 0$ and a system possessing only isotropic exchange interactions results in free boundary conditions for the magnetization, 
\begin{equation}
\frac{\partial \mathbf{m}}{\partial \mathbf{n}} = \mathbf{0}.
\end{equation}
However, the presence of the Dzyaloshinskii-Moriya interaction leads to a nontrivial pinning condition for the boundary magnetization with a chiral form that reflects its origin. For the form given in Eq. 1 of the article, the resulting boundary conditions are
\begin{eqnarray}
D m_z + 2 A \frac{\partial m_x}{\partial x} &= 0; \\
-D m_x + 2 A \frac{\partial m_z}{\partial x} &= 0; \\
D m_z + 2 A \frac{\partial m_y}{\partial y} &= 0; \\
-D m_y + 2 A \frac{\partial m_z}{\partial y} &= 0,
\end{eqnarray}
with all other spatial derivatives in $m$ vanishing. This reveals the chiral nature of the interaction, since the conditions couple the perpendicular magnetization component $m_z$ to its transverse components $m_{x,y}$.

\subsection{Perturbation theory of domain-wall eigenmodes with the DM interaction}

We consider a N{\'e}el wall configuration described by the normalized magnetization field,
\begin{equation}
\mathbf{m} = \left( \cos(\phi) \sin(\theta), \sin(\phi) \sin(\theta), \cos(\theta) \right),
\end{equation}
where the equilibrium state $(\theta_0,\phi_0)$ is given by
\begin{align}
\theta_0(\mathbf{r}) &= 2 \tan^{-1}\left( \exp\left[{y/\lambda} \right] \right), \\
\phi_0(\mathbf{r}) &= -\textrm{sgn}(D)\frac{\pi}{2},
\end{align}
and $\lambda = \sqrt{A/K_0}$ is the domain-wall width parameter. The sign of the Dzyaloshinskii-Moriya constant $D$ determines the chirality of the wall. To determine the spin-wave spectrum, we consider local fluctuations $\delta \mathbf{m}$ about this equilibrium ground state by applying a local gauge transformation such that the equilibrium magnetization is oriented along the local $z$ axis everywhere. As such, the fluctuations can be described by small variations in the local transverse components $\delta m_x$ and $\delta m_y$. By linearizing the Landau-Lifshitz-Gilbert equation of motion, we obtain the dynamical matrix equation for the fluctuations $\delta m_x$ and $\delta m_y$,
\begin{widetext}
\begin{equation}
\frac{\partial}{\partial t}\begin{pmatrix}
\delta m_x\\
\delta m_y\\
\end{pmatrix}=\frac{2\gamma_0 K_0}{M_s}
\begin{pmatrix}
\frac{D}{K_0} \sech(y/\lambda)\frac{\partial}{\partial x} & -V_{PT}\\
V_{PT}-\frac{D}{K_0\lambda} \sech(y/\lambda) & \frac{D}{K_0} \sech(y/\lambda)\frac{\partial}{\partial x}\\
\end{pmatrix}
\begin{pmatrix}
\delta m_x\\
\delta m_y
\end{pmatrix}
\label{eq:linearLLG}
\end{equation}
\end{widetext}
where $V_{PT}=\left(1-2\sech^2(y/\lambda)-\lambda^2 \nabla^2\right)$. In the absence of the Dzyaloshinskii-Moriya interaction, $D=0$, the dynamical matrix is described by a Schr{\"o}dinger-like equation with a P{\"o}schl-Teller potential. In this case, it has been shown by Winter~\cite{Winter:1961hw} for Bloch domain walls that the eigenfunctions are given by a mode localized to the wall along the $y$ direction,
\begin{equation}
\langle x,y| \xi_{k_x,0}  \rangle = \exp({i k_x x}) \textrm{sech}\left( \frac{y}{\lambda}\right),
\end{equation}
and propagating states of the form,
\begin{equation}
\langle x,y| \xi_{k_x,k_y}  \rangle = \exp({i \mathbf{k}_{||} \cdot \mathbf{x}_{||}  }) \left(  \textrm{tanh}\left( \frac{y}{\lambda}\right) -  i k_y \lambda \right).
\end{equation}
The unperturbed modes are described by the dispersion relation $\omega_k=\nu(1 + \lambda^2 |k_{||}|^2)$, where $k_{||}$ is the wavenumber parallel to the film plane.

In the presence of the Dzyaloshinskii-Moriya interaction, $D \neq 0$, the Bloch wall eigenfunctions are no longer solutions to Eq.~\ref{eq:linearLLG}. Nevertheless, we can obtain estimates of the changes to the eigenmode frequencies by using perturbation theory. The terms to be treated as perturbations are $\nu D \; \textrm{sech}(y/\lambda)\frac{\partial}{\partial x}$ and $ -\nu \frac{D}{\lambda} \; \textrm{sech}(y/\lambda)$ and need to be addressed within the space formed by the complete set of the unperturbed eigenfunctions $|\xi_{k_{\alpha},k_{\beta}}\rangle$. This is a degenerate space because of the quadratic dependence of the eigenvalues $\omega_k$ on  $k_{||}$. The matrix elements in space representation correspond to the integrals
\begin{widetext}
\begin{equation}
I_1=\int\int\,dx\,dy\; \xi_{k_{\alpha},k_{\beta}}^*(x,y)\, \left(\nu D \; \textrm{sech}(y/\lambda)\frac{\partial}{\partial x} \right)\xi_{k_{\alpha},k_{\beta}}(x,y),
\end{equation}
\begin{equation}
I_2=-\int\int\,dx\,dy\; \xi_{k_{\alpha},k_{\beta}}^*(x,y)\,\nu \frac{D}{\lambda} \; \textrm{sech}(y/\lambda)\,\xi_{k_{\alpha},k_{\beta}}(x,y),
\end{equation}
with
\begin{equation}
\xi_{k_{\alpha},k_{\beta}}(x,y)=A(\omega_k)\exp[i(k_\alpha x+k_\beta y)]\,\left(\tanh(y/\lambda)+i k_\beta\lambda\right),
\end{equation}
\end{widetext}
where $A(\omega_k)$ is a normalization constant and $k_\alpha$ ($k_\beta$) is the wave vector propagating in the positive or negative direction of the $x$ ($y$) axis spanning the complete degenerate space. The integrals yield
\begin{equation}
I_1=\frac{i \pi \gamma }{2 Ms}\left(\dfrac{1+2 k_y^2\lambda^2}{1+k_y^2\lambda^2}\right)D k_x (1\pm\sech(\pi k_y\lambda)),
\end{equation}
\begin{equation}
I_2=-\frac{ \pi \gamma }{2 Ms}\left(\dfrac{1+2 k_y^2\lambda^2}{1+k_y^2\lambda^2}\right)\frac{D}{\lambda}  (1\pm\sech(\pi k_y\lambda)).
\end{equation}
Considering these corrections the complete dispersion relation up to first order perturbation theory is 
\begin{equation}\label{eq:drneelw}
\Omega(k_x,k_y)=\frac{2\gamma K_o}{Ms}\left[\mp(k_x\lambda)\omega_k^1+\sqrt{\omega_k(\omega_k-\omega_k^1)}\right],
\end{equation}
with
\begin{equation}
\omega_k^1=\frac{\pi D }{4 K_o \lambda}\left(\dfrac{1+2 k_y^2\lambda^2}{1+k_y^2\lambda^2}\right) (1\pm\sech(\pi k_y\lambda)).
\end{equation}
The first term on the right hand side of eq (\ref{eq:drneelw}) is responsible for the nonreciprocity and the $\mp$ sign indicates the chirality and corresponds to considering a positive or a negative $D$ constant.

\bibliography{articles}

\end{document}